\documentstyle[aps,twocolumn,prl,epsf]{revtex}

\begin{document}

\newcommand{\be}{\begin{equation}}
\newcommand{\ee}{\end{equation}}
\newcommand{\bn}{\begin{eqnarray}}
\newcommand{\en}{\end{eqnarray}}
\def\beq{\begin{eqnarray}}
\def\eeq{\end{eqnarray}}
\def\lsim{\:\raisebox{-0.5ex}{$\stackrel{\textstyle<}{\sim}$}\:}
\def\gsim{\:\raisebox{-0.5ex}{$\stackrel{\textstyle>}{\sim}$}\:}
\def\d{\displaystyle}
\def\u{\underbar}

\draft

\twocolumn[\hsize\textwidth\columnwidth\hsize\csname @twocolumnfalse\endcsname

\title{\large \bf $VO_2$: a two-fluid incoherent metal?}

\author{M. S. Laad$^{1}$, L. Craco$^{2}$ and E. M\"uller-Hartmann$^{2}$}

\address{
$^{1}$Department of Physics, Loughborough University, LE11 3TU, UK \\
$^{2}$Institut f\"ur Theoretische Physik, Universit\"at zu K\"oln, 
77 Z\"ulpicher Strasse, D-50937 K\"oln, Germany \\}
\date{\today}
\maketitle

\widetext

\begin{abstract}
We present {\it ab initio} LDA+DMFT results for the many-particle density 
of states of $VO_{2}$ on the metallic side of the strongly first-order 
($T$-driven) insulator-metal transition.  In strong contrast to LDA 
predictions, there is {\it no} remnant of even correlated Fermi liquid 
behavior in the correlated metal. Excellent quantitative agreement with 
published photoemission and X-ray absorption experiments is found in the 
metallic phase.  We argue that the absence of FL-quasiparticles provides a 
natural explanation for the bad-metallic transport for $T > 340~K$.  Based 
on this agreement, we propose that the I-M transition in $VO_{2}$ is an 
orbital-selective Mott transition, and point out the relevance of orbital 
resolved one-electron and optical spectroscopy to resolve this outstanding 
issue. 
\end{abstract}   


]

\narrowtext

Metal-insulator transitions (MIT) in early-transition metal oxides (TMO)
are classic examples of correlation-driven Mott-Hubbard transitions~\cite{[1]}.
Strong multi-orbital correlations in the real crystal structure (RCS) have 
turned out to be indispensable for a quantitative description of the MIT in
these cases~\cite{[2]}.

A detailed understanding of the above cases is still somewhat elusive.  In
particular, it is still unclear whether the MIT is orbital selective, i.e., 
whether different orbital-resolved densities-of-states (DOS) are gapped 
(driving the Mott insulating state) at different values of $U, U'$ (defined 
below) or whether there is a single (common) Mott transition at a critical 
interaction strength~\cite{[3],[4]}.  Conflicting results, even for the same 
system, and within the same ($d=\infty$) approximation~\cite{[4]} have been 
obtained, necessitating more work to resolve this issue.

In this communication, we address precisely this issue in vanadium dioxide
($VO_{2}$).  Using the state-of-the-art LDA+DMFT~\cite{[5]} technique, we 
first show how excellent quantitative agreement with the full, local
 one-electron spectral function is achieved using LDA+DMFT(IPT).  We then 
build on this agreement, claiming that the MIT in $VO_{2}$ is orbital 
selective, and represents a concrete, {\it ab initio} realization of the 
two-fluid scenario for MIT in strongly correlated systems.

$VO_{2}$ shows a spectacular MIT at $T=340~K$ from a low-$T$ monoclinic, Mott
insulating phase with spin dimerization along the crystallographic $c$-axis to
a local moment paramagnetic metallic phase.  This $T$-driven MIT has attracted
intense scrutiny: because of the additional complication of the 
antiferroelectric displacement of the $VO_{6}$ octahedra in a correlated 
system, an unambiguous characterization of the MIT (relative importance of 
Mott-Hubbard versus Peierls dimerization) is somewhat difficult.  Various 
observations have been cited in support of both scenarios~\cite{[1]}, and 
theoretical models detailing the importance of these effects have been 
proposed~\cite{[6],[7]}.

One-electron spectroscopies constitute a reliable fingerprint of the changes
in the single-particle spectral

\begin{figure}[htb]
\epsfxsize=3.5in
\epsffile{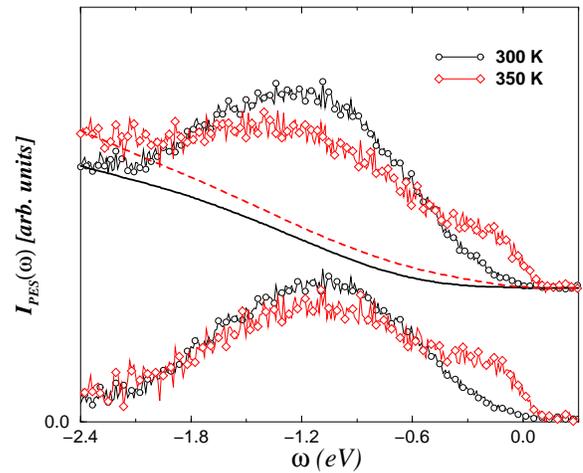}
\caption{Photoemission spectrum of $VO_{2}$ across the metal-insulator 
transition, from~\protect\cite{[11]}. Top: raw data; bottom: with the 
$O$-2p background subtracted in a way indicated by the region below the 
solid~$(300~K)$ and dashed~$(350~K)$ lines.}
\label{fig1}
\end{figure}

\hspace{-0.3cm}function (and hence those in the 
{\it correlated} electronic structures) across a phase transition.  
Many papers have studied the MIT in $VO_{2}$ by photoelectron- and X-ray 
absorption spectroscopy (PES and XAS)~\cite{[8],[9],[10]}.  A perusal of the 
available literature requires care to pinpoint robust features (weakly 
dependent on incident photon energy); in what follows, we use only those 
published data which satisfy this criterion in order to make comparison with
theory.  Usually, the experiments have been carried out with low incident 
photon energies ($50-70$~eV) except in Ref.~\cite{[11]} ($1200$~eV).  
However, the main features, namely: (i) a broad step-like feature crossing 
$E_{F}$ in the $R$ phase, (ii) the ``lower Hubbard band'' centered at 
$-1$~eV, and (iii) the huge $O$ 2p-band contribution from $-2$~eV, seem to 
be roughly consistent in all these measurements.  Upon subtraction of the 
$O$-2p contribution, the corresponding ``Vanadium'' $t_{2g}$ DOS shows 
noticeably more spectral weight at lower energies relative to that at 
$-1$~eV (Fig.~\ref{fig1}). In this context, great care has to be taken, as, 
in view of the dynamical transfer of spectral weight over large energy 
scales characteristic of a correlated system, it is not a priori clear 
whether the subtracted ``background'' contains significant intrinsic strong 
correlation (continuum) contributions.  Additional subtraction due to surface 
contributions to the incoherent part~\cite{[11]} results in noticeably more 
spectral weight at low energy compared to that at $-1.5~eV$.  

With these caveats, a consistent picture of the changes in the actual 
electronic structure across the MIT seems to emerge.  First, the transition is
clearly driven by the dynamical transfer of spectral weight from high to low
energy over a large energy scale of $O(2~eV)$, characteristic of a Mott-Hubbard
scenario.  Second, in the metallic ($R$) phase, there is no semblance of a 
Fermi liquid-like peak at low energy: rather, the low-energy feature is 
more-or-less incoherent.  This is in good accord with the high resistivity 
in the (metallic) $R$ phase, as also with the {\it linear}-in-$T$ dependence 
without saturation upto $900~K$~\cite{[12]}. It is natural to postulate a 
common electronic scattering mechanism leading to such observations.  In the 
insulating ($I$) phase, clear evidence of insulating gap formation is visible 
in PES.  From Fig.~\ref{fig3} of Ref.~\cite{[11]}, we estimate the gap 
value $E_{G}\simeq 0.2-0.3~eV$, consistent with that extracted from optical
data~\cite{[13]}.  

The XAS data probe the unoccupied part of the full spectral function, and 
show a number of interesting features in $VO_{2}$.  First, we notice that the
XAS intensity is much larger (for $0\le\omega\le 2.0$~eV) than the XPS 
intensity (for $-2.0\le\omega\le 0$~eV).  Second, the broad low-energy 
``step-like'' feature seems actually to be the lower part of the intense 
feature in XAS pulled below $E_{F}$, and is {\it not} a thermally smeared 
correlated Fermi liquid resonance.  As mentioned above, this fits in nicely 
with the transport data; in any case,  we do not expect a quasicoherent FL 
scale to survive beyond very low temperatures in a correlated system close 
to a MI transition.

Beginning with the LDA bandstructure of $VO_{2}$ in the monoclinic crystal 
structure corresponding to the insulating phase~\cite{[14]}, with subtraction 
of terms to avoid double-counting of interactions treated on average by LDA, 
the one-electron Hamiltonian is,

\be
H_{LDA}^{(0)}=\sum_{{\bf k}\alpha\beta\sigma}\epsilon_{\alpha\beta}
({\bf k})c_{{\bf k}\alpha\sigma}^{\dag}c_{{\bf k}\beta\sigma} 
+ \sum_{i\alpha\sigma}\epsilon_{i\alpha\sigma}^{0}n_{i\alpha\sigma},
\ee
where 
$\epsilon_{i\alpha\sigma}^{0}=\epsilon_{i\alpha\sigma}-U(n_{\alpha\bar\sigma}
-\frac{1}{2})+\frac{1}{2} \sigma J_{H}(n_{\alpha \sigma}-1)$.  
Here, $U,U',J_{H}$ are the Coulomb interactions in the $t_{2g}$ shell, and 
are defined below.  The lowest lying $xy$ (in the $M$ notation) band is the 
most heavily populated, while the $yz\pm zx$ bands are less populated.  The 
bonding-antibonding splitting~\cite{[14]} is clearly visible in the results, 
which also show that the total bandwidth is about $W \simeq 2.3~eV$, contrary 
to much smaller previous~\cite{[7]} estimates based on model calculations. 
Sophisticated LDA calculations do yield a miniscule, almost zero-gap, Peierls 
insulator~\cite{[6]}, but obviously do not reproduce the Mott insulator.  As 
emphasized by Zylbersteyn {\it et al.}~\cite{[7]}, it is inconceivable that 
the antiferroelectric distortion alone could open up a charge gap 
($E_{G}\simeq 0.6~eV$), requiring proper incorporation of correlation effects. 
    
Including multi-orbital correlations, the full Hamiltonian is

\bn
\nonumber
H &=& H_{LDA}^{(0)}+U\sum_{i\alpha}n_{i\alpha\uparrow}n_{i\alpha\downarrow} 
+ \sum_{i\alpha\beta\sigma\sigma'} U_{\alpha\beta}
n_{i\alpha\sigma}n_{i\beta\sigma'} \\
&-& J_{H}\sum_{i\alpha\beta}{\bf S}_{i\alpha} \cdot{\bf S}_{i\beta}.
\en

This multi-orbital Hubbard model is solved using LDA+DMFT(IPT)~\cite{[3]}.
Accurate low- and high energy behavior is ensured by requiring strict 
numerical compliance with the Friedel-Luttinger sum rule (low-energy) and 
the first three moments of the DOS (high energy).  In our approach, the MI 
transition is driven by the large change in the dynamical spectral weight 
transfer accompanying small changes in the renormalized antiferroelectric 
displacement in the {\it correlated} picture.  Hence the transition is of 
the Mott-Hubbard type, but the antiferroelectric coupling also plays an 
important role: in fact, it acts like an ``external field'' in the orbital 
($t_{2g}$) sector.  Changes in this external field drives changes in orbital 
occupation on the one hand, and results in a lowering of the free energy (of 
the metallic solution) as a function of $\Delta_{\alpha\beta}$, stabilizing 
the metallic solution at a critical value of $\Delta_{\alpha\beta}$, as found 
in the calculations. Comparison with data for the insulating state requires 
incorporation of the $c$-axis dimerization~\cite{[7]} into our DMFT procedure,
and requires extension of the present approach.  Hence, we focus only on the 
metallic state in what follows.
 
We now compare our theoretical results to experiment, showing how 
excellent quantitative agreement with the {\it full} one-electron
spectrum in the metallic ($R$) phase is achieved in this picture.  The
calculations were done at $T_{1}=150~K < T_{MI}$ (I-phase) and 
$T_{2}=370~K > T_{MI}$ (M-phase). We used parameter values identical to 
those used in Ref.~\cite{[3]}: the LDA bandwidth $W\simeq 2.0~eV$, and 
interactions specified by $U\simeq 5.0~eV, J_{H}=1.0~eV$ and 
$U'\simeq (U-2J_{H})=3.0~eV$ for the I-phase.  To account for reduction of 
the effective interaction due to dynamical (selfconsistent) screening 
processes relevant upon the transition to the M-phase, we use reduced 
values of $U\simeq 4.6~eV, U'\simeq 2.6~eV$ in the $M$-phase, while $J_{H}$, 
being almost unscreened, is kept fixed.  As described in Ref.~\cite{[3]}, 
the MIT in our picture results from large changes in dynamical transfer of 
spectral weight from high- to low energies following small changes in the 
{\it renormalized} antiferroelectric distortion in $VO_{2}$.  

\begin{figure}[htb]
\epsfxsize=3.5in
\epsffile{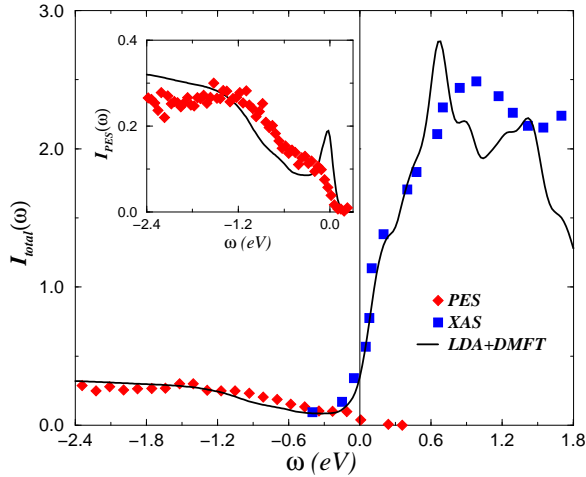}
\caption{Comparison of theoretical (LDA +DMFT) result for the total 
one-electron spectral function in the metallic phase of $VO_{2}$ to the 
experimental results taken from~\protect\cite{[11]} (for PES) and 
from~\protect\cite{[9]} (for XAS). The inset shows a blow-up of the PES 
spectrum.  The theoretical result is compared with the raw data without 
subtraction of the $O$-2p and surface contribution. Subtraction of the 
surface contribution as in~\protect\cite{[11]} would lead to even better 
agreement, as explained in the text.}  
\label{fig2}
\end{figure}

\begin{figure}[htb]
\epsfxsize=3.5in
\epsffile{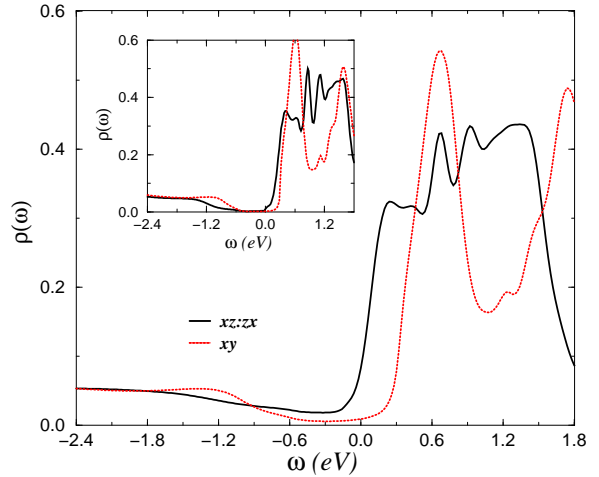}
\caption{Orbital-resolved one electron spectral functions for the metallic 
phase of $VO_{2}$ (main frame) and for the insulating phase without 
spin dimerization (inset).  Only the $d_{yz\pm zx}$ orbital DOS crosses 
$E_{F}$ in the metallic phase; the $d_{xy}$ DOS still shows Mott-Hubbard 
insulating features, showing the ``two-fluid'' character of the MIT in 
$VO_{2}$.}
\label{fig3}
\end{figure}

In Fig.~\ref{fig2}, we show the theory-experiment comparison for the 
total (photoemission + X-ray absorption) intensity, $I_{total}(\omega)$ for 
the metallic ($R$) phase. Remarkably,

\hspace{-0.3cm}the agreement between theory and 
experiment in the region $-2.4\le\omega\le 0.8~eV$ is excellent.  We have 
not tried to fit the data for $\omega < -2.4~eV$ and $\omega > 0.8~eV$ 
because only part of the $O$-2p and $e_{g}$ bands of $V$ projected onto the 
$t_{2g}$ states have been included 
in our LDA+DMFT calculation. We emphasize that even better agreement could 
have been obtained if the subtraction of the surface contributions would 
have been carried out without taking out the $O$-2p ``background'' in 
Fig.~\ref{fig1}. This would have had the effect of increasing the weight 
in the region $-0.3\le\omega\le 0~eV$ and decreasing the weight in the 
region $-1.8\le\omega\le -0.5~eV$~\cite{[11]}, further improving the fit.

Several features of the experimental spectra are readily interpreted in the 
light of the theoretical calculation.  First, most of the low-energy spectral
weight actually corresponds to (largely incoherent) electronic states close to 
the minimum of the pseudogap in the spectral function.  There is no indication
of even a strongly renormalized FL-like resonance.  The DOS at $E_{F}=0~eV$
is drastically reduced compared to the LDA value~\cite{[14]}, in excellent 
agreement with the data, and in full accord with the bad metallic resistivity
in the $R$ phase.  The lower Hubbard band feature around $-1.2~eV$ is
reproduced very well, as is the shoulder-like feature at $0.3~eV$ in XAS.
In~\cite{[3]}, we found excellent agreement with the optical effective mass,
$m^{*}/m_{b} \simeq 3.0$, a value that is also consistent with the observed 
enhancement of the paramagnetic spin susceptibility in the $R$ phase.  Thus,
these results represent a quantitatively accurate description of basic 
spectroscopic and thermodynamic results for the metallic state.

Armed with this agreement, we proceed to analyze the MIT in $VO_{2}$ in
more microscopic detail. Fig.~\ref{fig3} shows the orbital-resolved DOS for 
the metallic phase.  Clearly, only the $yz\pm zx$ DOS gives rise to the 
metallic behavior; the $xy$ orbital DOS is still gapped at a value close 
to that for the (undimerized, but this is {\it not} the real lowest state 
in the $I$ phase) paramagnetic insulating state.  This observation, 
constituting an explicit manifestation of the two-fluid scenario for the 
Mott transition in $VO_{2}$ driven by multi-orbital correlations, is the 
central result of our analysis. If confirmed by experiment, this result 
would imply concrete evidence of a multi-orbital correlation driven MIT in 
$VO_{2}$.  We suggest that orbital resolved electron spectroscopies can 
resolve this general issue of importance for transition-metal oxides. Based 
on the above analysis, we expect that only the $yz\pm zx$ DOS will cross 
$E_{F}$ as $T$ is increased above $T_{MI}$, while the $xy$ DOS will still 
show insulating behavior.  A detailed analysis of the anisotropic optical
spectra across the MI transition driven by strong changes in orbital 
occupation can also verify the proposed scenario~\cite{[Tobe]}. 

The above results are also consistent with the large (bad-metallic), 
linear-in-$T$ resistivity observed in the high-$T$ metallic phase of $VO_{2}$.
Indeed, within our DMFT treatment of local (multi-orbital) electronic
correlations, the resistivity is expected to show a quadratic temperature 
dependence for $T < T_{coh}$, a lattice FL coherence scale, but a 
linear-in-$T$ dependence above $T_{coh}$ up to high temperatures 
$T^{*} \simeq J \simeq 4t^{2}/U\simeq 0.1~eV$~\cite{[15]}.  For large 
$U/t$, and near the MIT, the resistivity increases way beyond the Mott 
limit, and increases without saturation. In $VO_{2}$, the MIT occurs at 
$T_{MI}=340~K$, which is much higher than an effective FL coherence scale 
(which might have been visible if $T_{MI}$ could have been low enough).  
Hence, the low-$T$ FL-like resistivity is not observed, while the 
linear-in-$T$ resistivity up to $T^{*}=900~K$ without any hint of saturation 
can now be identified with the incoherent local (spin-orbital) moment regime 
of the metallic phase in a multi-band Hubbard model.  Using this analogy, we 
also see that the optical conductivity will be almost completely incoherent in
the $M$ phase.

We have not considered the insulating phase with spin dimerization in this 
communication.  To do so requires an explicit consideration of the dynamical 
effects associated with the $c$-axis dimerization.  This requires a non-trivial
extension of DMFT, and we leave it for the future.

In conclusion, we have shown how the two-fluid picture of the Mott transition 
in $VO_{2}$ based on the LDA+DMFT(IPT) yields excellent quantitative 
agreement with the full one-electron spectrum in the metallic phase.  
Several features of the bad-metallic transport are also understandable in 
this picture.  Orbital selective one-electron- and optical spectroscopy can 
confirm/refute this outstanding issue that, in the final analysis, is a 
manifestation of Mott transitions driven by strong, multi-orbital electronic 
correlations. 

\acknowledgments
Work carried out (LC) under the auspices of the Sonderforschungsbereich 
608 of the Deutsche Forschungsgemeinschaft.

\end{document}